\documentclass[11pt,a4paper,english,superscriptaddress]{revtex4}

\usepackage{amssymb}
\usepackage{amsmath}
\usepackage{graphicx}
\usepackage{babel}
\usepackage[T1]{fontenc}
\usepackage{hyperref}

\begin{document}
\title{On the superfield effective potential in three dimensions}

\author{A. F. Ferrari}
\email{alysson.ferrari@ufabc.edu.br}
\affiliation{Centro de Ci\^encias Naturais e Humanas, Universidade Federal do ABC, Rua Santa Ad\'elia, 166, 09210-170, Santo Andr\'e, SP, Brazil}

\author{M. Gomes}
\email{mgomes@fma.if.usp.br}
\affiliation{Instituto de F\'\i sica, Universidade de S\~ao Paulo\\
Caixa Postal 66318, 05315-970, S\~ao Paulo, SP, Brazil}

\author{A. C. Lehum}
\email{lehum@fma.if.usp.br}
\affiliation{Instituto de F\'\i sica, Universidade de S\~ao Paulo\\
Caixa Postal 66318, 05315-970, S\~ao Paulo, SP, Brazil}
\affiliation{Escola de Ci\^encias e Tecnologia, Universidade Federal do Rio Grande do Norte\\
Caixa Postal 1524, 59072-970, Natal, RN, Brazil}

\author{J. R. Nascimento} 
\email{jroberto@fisica.ufpb.br}
\affiliation{Departamento de F\'{\i}sica, Universidade Federal da Para\'{\i}ba\\
 Caixa Postal 5008, 58051-970, Jo\~ao Pessoa, Para\'{\i}ba, Brazil}

\author{A. Yu. Petrov}
\email{petrov@fisica.ufpb.br}
\affiliation{Departamento de F\'{\i}sica, Universidade Federal da Para\'{\i}ba\\
 Caixa Postal 5008, 58051-970, Jo\~ao Pessoa, Para\'{\i}ba, Brazil}

\author{E. O. Silva}
\email{edilberto@fisica.ufpb.br}
\affiliation{Departamento de F\'{\i}sica, Universidade Federal da Para\'{\i}ba\\
Caixa Postal 5008, 58051-970, Jo\~ao Pessoa, Para\'{\i}ba, Brazil}

\author{A. J. da Silva}
\email{ajsilva@fma.if.usp.br}
\affiliation{Instituto de F\'\i sica, Universidade de S\~ao Paulo\\
Caixa Postal 66318, 05315-970, S\~ao Paulo, SP, Brazil}

\begin{abstract}
We develop the superfield approach to the effective potential in three dimensions and calculate the one-loop and two-loop K\"{a}hlerian effective potential in commutative and noncommutative cases.
\end{abstract}

\maketitle

The effective potential is a key object in quantum field theory whose study allows to obtain fundamental information about different aspects of an arbitrary physical theory, such as the structure of the vaccum, spontaneous symmetry breaking, phase transitions, etc.~\cite{CW}. In the context of recent investigations of spontaneous supersymmetry breaking~\cite{Sei}, the study of the effective potential in supersymmetric field theories seems to be highly relevant. At the same time, up to now, the supersymmetric effective potential was well studied only in four-dimensional space-time, where the whole methodology for its evaluation, based on the use of the superfield approach on all steps of the calculations, was developed~\cite{BK0,WZ}. Such methodology was applied for different supersymmetric field theories, such as the Wess-Zumino model, both in commutative~\cite{WZ} and noncommutative cases~\cite{WZNC}, general chiral superfield model~\cite{GC} and super-Yang-Mills theory~\cite{SYM}. The superfield approach to the study of the supersymmetric effective potential in three space-time dimensions was not well developed despite a number of interesting results regarding three-dimensional supersymmetric field theories, especially noncommutative ones~\cite{3d}. 

Therefore a natural problem is the development of a manifestly supercovariant methodology for the calculation of the effective potential in a three-dimensional supersymmetric field theory. Some difficulties related to this subject were pointed out in~\cite{Burgess,Lehum}. 
One important point is that in writing the vacuum expectation value of a scalar superfield $\Phi(x,\theta)=A(x)+\theta^\alpha \psi_\alpha - \theta^2 F(x)$, we would have in general
\begin{equation}
\label{motiv1}
<\Phi(x,\theta)>=a - \theta^2 f\,,
\end{equation}

\noindent
with $a$ and $f$ constants ($<\psi_\alpha>=0$ to preserve Lorentz invariance). If we allow $f \neq 0$, the background-dependent propagator for the quantum field $\Phi$ becomes non-local in the $\theta$ variable, making cumbersome the calculation of supergraphs. On the other hand, if $f = 0$, the background superfield $<\Phi>$ would be independent of the grassmanian coordinate of the superspace and, as a consequence, every superspace integral of a polinomial of background superfield would identically vanish. In four spacetime dimensions, such difficulties can be surmounted, for example, by using the methodology developed in~\cite{BK0,WZ} for the evaluation of the superfield effective potential. However, the structure of three-dimensional supersymmetric models differs in relevant aspects when compared to the four-dimensional theories. In particular, in three-dimensions there are neither chiral nor anti-chiral superfields, which play a fundamental role in the approach of~\cite{BK0,WZ}. 
In this work, we will show how the above mentioned method must be modified for three-dimensional theories. We shall work on a noncommutative spacetime, but our method also can be applied in the commutative case. For simplicity, we will restrict ourselves to the case of a theory involving a single scalar superfield. 

We start with the following three-dimensional superfield theory which is described by a general scalar superfield action (see f.e.~\cite{SGRS}):
\begin{eqnarray}
\label{sfa}
S[\Phi]=\int d^5z \left[\frac{1}{2}\Phi D^2\Phi-V(\Phi)\right],
\end{eqnarray}

\noindent
where $\Phi$ is a scalar superfield. We will start by evaluating the superfield effective action, in loop expansion~\cite{BO}. To do it, we make a shift in the field $\Phi$,
\begin{eqnarray}
\Phi\to\Phi_0+\sqrt{\hbar} \, \phi,
\end{eqnarray}

\noindent
where $\Phi_0$ is a background (super)field (further we omit the index 0), and $\phi$ is a quantum one, which is contracted into propagators.
As a result, the classical action (\ref{sfa}) takes the form
\begin{eqnarray}
\label{2}
S[\Phi,\phi]&=&S[\Phi]+\int d^5z\left(\hbar\frac{1}{2}\phi[D^2-V^{\prime\prime}(\Phi)]\phi-\hbar^{3/2}\frac{1}{3!}V^{\prime\prime\prime}(\Phi)\phi^{3}_{*}-
\hbar^2\frac{1}{4!}V^{(IV)}(\Phi)\phi^{4}_{*}
\right)\nonumber\\
&&+\ldots,
\end{eqnarray}

\noindent
where dots are for irrelevant terms in the two-loop approximation. Here the star symbol denotes the fact that the usual product of the fields is replaced by a Moyal-Groenewold one. The linear terms in $\phi$ are omitted since they produce only one-particle-reducible contributions which are irrelevant in the context of the effective action.
The effective action $\Gamma[ \Phi]$ is defined by the expression (see more details in~\cite{Jackiw,BO})
\begin{eqnarray}
\exp\left(\frac{i}{\hbar}\Gamma[\Phi]\right)\,= \mathcal{N} \,\int D\phi \, \exp\left(\frac{i}{\hbar}S[\Phi,\phi]\right),
\end{eqnarray}

\noindent
where $\mathcal{N}$ is a normalization factor. The general structure of the effective action can be cast in a form similar to the four-dimensional case~\cite{BK0,WZ}:
\begin{eqnarray}
\Gamma[\Phi]=\int d^5z \, K(\Phi)+\int d^5z \, F(D_{\alpha}\Phi,D^2\Phi;\Phi),
\end{eqnarray}

\noindent
where the $K(\Phi)$ is the K\"{a}hlerian effective potential and depends only on the superfield $\Phi$ but not on its derivatives, and $F$ is called auxiliary fields effective potential whose key property is its vanishing in the case when all derivatives of the superfields are equal to zero. It is easy to see that $F$ is at least of the second order in the auxiliary field of the scalar supermultiplet. 
It can be explicitly written as
\begin{eqnarray}
F(D_{\alpha}\Phi,D^2\Phi;\Phi)\,=\,F_{2.1}(\Phi)D^2\Phi+F_{2.2}(\Phi)D^{\alpha}\Phi D_{\alpha}\Phi+\ldots,
\end{eqnarray}

\noindent
where the $F_{2.1}(\Phi)$ and $F_{2.2}(\Phi)$ are functions of $\Phi$ only but not of its derivatives, and the dots correspond to terms with four or more supercovariant derivatives. It is clear that this approach does not require to impose the condition $D_{\alpha}\Phi=0$ which is known to imply in difficulties in the interpretation of the results (see f.e. \cite{Burgess}). 

We will work with a loop expansion for the effective action $\Gamma$, 
\begin{eqnarray}
\label{4}
\Gamma[ \Phi]=S[ \Phi]+\hbar \Gamma^{(1)}[ \Phi]+\hbar^2\Gamma^{(2)}[ \Phi]+\ldots,
\end{eqnarray}

\noindent
the K\"ahlerian potential $K$,
\begin{eqnarray}
K(\Phi)=V(\Phi)+\sum_{L=1}^{\infty}\hbar^LK_L(\Phi),
\end{eqnarray}

\noindent
and similarly for $F$. 

We start by considering the one-loop effective action in the form
\begin{eqnarray}
\label{g1}
\Gamma^{(1)}=\frac{i}{2}{\rm Tr}\ln[D^2-V^{\prime\prime}(\Phi)] + c\,,
\end{eqnarray}

\noindent
where $c$ is a constant coming from the normalization of the effective action. 
The more convenient normalization is $\Gamma[0] = 0$, which
corresponds to $c = \frac{i}{2}{\rm Tr}\ln(D^2)$. 

As a first approximation, let us consider the K\"{a}hlerian effective action. From a formal viewpoint this corresponds to disregarding all terms depending on derivatives of $\Phi$ (both common and spinor ones), and allows us to calculate the quantum corrections to $V(\Phi)$. In this case, we can write
\begin{eqnarray}
\label{g12}
\Gamma^{(1)}=\frac{i}{2}{\rm Tr}\ln[\Box-V^{\prime\prime}(\Phi)D^2].
\end{eqnarray}

\noindent
This expression can be represented via the Schwinger proper-time representation \cite{Ojima,ourpt}:
\begin{eqnarray}
\Gamma^{(1)}&=&\frac{i}{2}{\rm Tr}\int_0^{\infty}\frac{ds}{s}e^{is[\Box-V^{\prime\prime}(\Phi)D^2]} \nonumber\\
&=&\frac{i}{2}\int d^5z\int_0^{\infty}\frac{ds}{s}e^{is[\Box-V^{\prime\prime}(\Phi)D^2]}\delta^5(z-z')|_{z=z'}\,.
\end{eqnarray}

\noindent
Again, since we are calculating only the K\"{a}hlerian part of the effective action, we have
\begin{eqnarray}
\label{gamma1}
\Gamma^{(1)}= \frac{i}{2} {\rm Tr}\int d^5z\int_0^{\infty}\frac{ds}{s}e^{-isV^{\prime\prime}(\Phi)D^2}e^{is\Box}\delta^5(z-z')|_{z=z'}\,,
\end{eqnarray}

\noindent
or, using that $(D^2)^2=\Box$, 
\begin{eqnarray}
\label{expa}
e^{-isV^{\prime\prime}(\Phi)D^2}=\sum_{n=0}^{\infty}\frac{\left[-isV^{\prime\prime}(\Phi)\right]^{2n+1}}{(2n+1)!}\Box^nD^2+\ldots.
\end{eqnarray}

\noindent
Here the dots stand for terms which do not contribute to the integral. 
At this point, we can clearly state the difference between the calculation of $\Gamma^{(1)}$ in four- and three-spacetime dimensions. In four dimensions~\cite{BK0,WZ}, $\Gamma^{(1)}$ is given by an expression similar to Eq.~(\ref{gamma1}), but there are more independent structures involving superderivatives and chiral and antichiral background superfields. The calculation of the exponential similar to Eq.~(\ref{expa}) involves the solving of a coupled set of differential equations, whose solutions can be found but are of rather cumbersome form. In three dimensions the number of independent structures is much smaller, actually only terms involving a $D^2$ will be relevant to the calculation of the K\"{a}hlerian effective action. We will shortly show that these terms can be directly summed, thus providing a closed-form expression for $\Gamma^{(1)}$.

Let us now consider a function $U(x,x';s)=e^{is\Box}\delta^3(x-x')$. Its key property is that
\begin{eqnarray}
i\frac{\partial U}{\partial s}=-\Box U,
\end{eqnarray}

\noindent
which allows us to obtain
\begin{eqnarray}
\Box^n U(x,x';s)|_{x=x'}\equiv \Box^ne^{is\Box}\delta^3(x-x')|_{x=x'}=\frac{\sqrt{i}}{8\pi^{3/2}}\left(-i\frac{d}{ds}\right)^n\frac{1}{s^{3/2}}=\frac{i^{n+1/2}}{8\pi^{3/2}}\frac{(2n+1)!!}{2^ns^{3/2+n}}.
\end{eqnarray}

\noindent
From Eq.~(\ref{expa}), after calculating the trace using that $D^2\delta^2(\theta-\theta')|_{z=z'}=1$ and $\frac{(2n+1)!!}{(2n+1)!}=\frac{1}{(2n)!!}=\frac{1}{2^nn!}$, we obtain
\begin{eqnarray}
\Gamma^{(1)}=\frac{i}{16\pi^{3/2}}
\int d^5z\int_0^{\infty}\frac{ds}{s}\sum_{n=0}^{\infty}\frac{\left[-\sqrt{-i}\,V^{\prime\prime}(\Phi)\right]^{2n+1}}{4^nn!}s^{n-1/2}.
\end{eqnarray}

\noindent
By performing the summation, we end up with
\begin{eqnarray}\label{eq1}
\Gamma^{(1)}=-\frac{i\sqrt{i}}{16\pi^{3/2}}\int d^5z V^{\prime\prime}(\Phi)\int_0^{\infty}\frac{ds}{s^{3/2}}e^{-is[\frac{(V^{\prime\prime}(\Phi))^2}{4}]}.
\end{eqnarray}

\noindent
After an appropriate analytic continuation, we recognize in (\ref{eq1}) the integral defining a Gamma function, and we finally arrive at
\begin{eqnarray}
\Gamma^{(1)}=\frac{1}{16\pi}\int d^5z \left[V^{\prime\prime}(\Phi)\right]^2.
\end{eqnarray}

\noindent
This is our final expression for the one-loop K\"ahlerian effective action. It is positively defined, as it should be in a supersymmetric theory. We note its finiteness, and we can also observe that this expression holds also in the noncommutative case. Indeed, since the background superfield is constant in the space-time, the Moyal-Groenewold product of these superfields reduces to the usual one (for reviews on noncommutative field theories defined by means of the Moyal-Groenewold product, see for example~\cite{reviews}).

Now, let us go to two loops. We will consider the noncommutative case, the reduction to the commutative one presents no difficulties. Applying the expansion given in Eqs.~(\ref{2}--\ref{4}), we can find that the expression for the two-loop effective action $\Gamma^{(2)}$ looks like,
\begin{eqnarray}
\Gamma^{(2)}=-i\int D\phi \exp\left(\frac{i}{2}\phi[D^2-V^{\prime\prime}(\Phi)]\phi\right)
\left[\frac{1}{2}\left(\frac{1}{3!}V^{\prime\prime\prime}(\Phi)\phi_{*}^{3}\right)^2-
\frac{1}{4!}V^{(IV)}(\Phi)\phi_{*}^{4}\right].
\end{eqnarray}

The two-loop contributions are given by two supergraphs,

%\Lengthunit=1mm
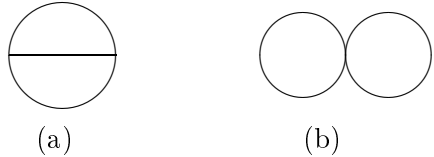
\begin{center}
\begin{picture}(200,60)
\put(30,30){\circle{40}}
\put(10,30){\line(1,0){40}}
\put(20,-5){(a)}
\put(120,30){\circle{30}}
\put(152,30){\circle{30}}
\put(120,-5){(b)}
\end{picture}
\end{center}

Since we are interested in calculating the two-loop contribution to the K\"ahlerian effective action, we can effectively assume that $D^\alpha \Phi = 0$, so that the background field dependent mass $M=V^{\prime\prime}(\Phi)$ is independent of $\theta$, thus the simple propagator
\begin{eqnarray}
<\phi(z_1)\phi(z_2)>=-i\frac{D^2+M}{\Box-M^2}\,\delta^5 (z_1 - z_2),
\end{eqnarray}
can be used. We also remind that the background superfield is constant, so it is not affected by the Moyal product~\cite{WZNC}. 

The vertices in the noncommutative case look like:
\begin{align}
\int d^5z \, & V^{\prime\prime\prime}(\Phi)\phi^{*3}\\
&=\int d^2\theta\int\frac{d^3k_1d^3k_2d^3k_3}{(2\pi)^9}(2\pi)^3\delta(k_1+k_2+k_3)\cos(k_1\wedge k_2)V^{\prime\prime\prime}(\Phi)\phi(k_1)\phi(k_2)\phi(k_3)\nonumber;\\
\int d^5z \, & V^{(IV)}(\Phi)\phi^{*4}\nonumber\\
=&\frac{1}{3}\int d^2\theta\int\frac{d^3k_1d^3k_2d^3k_3d^3k_4}{(2\pi)^{12}}(2\pi)^3\delta(k_1+k_2+k_3+k_4)\times
\nonumber\\
&\times
[\cos(k_1\wedge k_2)\cos(k_3\wedge k_4)+\cos(k_1\wedge k_3)\cos(k_2\wedge k_4)+\cos(k_1\wedge k_4)\cos(k_2\wedge k_3)]\times\nonumber\\
&\times V^{(IV)}(\Phi)
\phi(k_1)\phi(k_2)\phi(k_3)\phi(k_4).
\end{align}

\noindent
Here $k \wedge p = k^\mu \Theta_{\mu\nu} p^\nu$, where $ \Theta_{\mu\nu}$ is the matrix characterizing the underlying noncommutativity of the spacetime.

Thus, the contributions from diagram (a) and (b) respectively, after trivial D-algebra transformations, look like
\begin{eqnarray}
\Gamma^{(2)}_{a}&=&\frac{1}{8}\int d^5z\,
\left[V^{\prime\prime\prime}(\Phi)\right]^2M\int\frac{d^3kd^3l}{(2\pi)^6}\frac{1+\cos(2k\wedge l)}
{(k^2+M^2)(l^2+M^2)[(k+l)^2+M^2]},
\end{eqnarray}

\noindent
and
\begin{eqnarray}
\Gamma^{(2)}_{b}&=&-\frac{1}{12}\int d^5z \, V^{(IV)}(\Phi)\int\frac{d^3kd^3l}{(2\pi)^6}\frac{2+\cos(2k\wedge l)}
{(k^2+M^2)(l^2+M^2)}.
\end{eqnarray}

\noindent
The commutative result is obtained by setting the noncommutativity to zero before integrating these equations. In the noncommutative case, we calculate these integrals using well-known relations (see f.e. \cite{Alv}). Proceeding in a similar way to~\cite{WZNC}, and considering the noncommutativity matrix as $\Theta_{\mu\nu} = \varepsilon_{0\mu\nu} \Theta$, we obtain
\begin{eqnarray}
\label{g1a}
\Gamma_{a}^{(2)}&=&\frac{1}{8}\int d^5z\, \left[V^{\prime\prime\prime}(\Phi)\right]^2M
\left[
-\frac{1}{32\pi^2\epsilon}-\frac{1}{32\pi^2}\ln\frac{M^2}{\mu^2}+\frac{3}{256\pi^2\Theta^2M^4}\right] + \mathcal{O}(\Theta).
\end{eqnarray}

\noindent
The divergence can be canceled via an appropriate counterterm. Also, one can conclude that this contribution is singular at $\Theta\to 0$ which is a natural consequence of the fact that in the commutative limit this contribution is divergent. This singularity has the same nature as the common  UV/IR infrared singularity characteristic of noncommutative theories.

For the $\Gamma_{b}$, we proceed in a similar way and find
\begin{eqnarray}
\label{g2}
\Gamma_{b}^{(2)}&=&-\frac{1}{12}\int d^5z \, V^{(IV)}(\Phi)\left[\frac{M^2}{8\pi^2}+\frac{1}{64\pi^2\Theta^2M^2}
\right].
\end{eqnarray}
In this case, there is no UV divergence but there is again a $\Theta=0$ singularity. The whole two-loop K\"ahlerian effective action is hence a sum of (\ref{g1a}) and (\ref{g2}). 

In the commutative case these expressions look like
\begin{eqnarray}
\label{g1ac}
\Gamma_{a,{\rm C}}^{(2)}&=&\frac{1}{4}\int d^5z\, \left[V^{\prime\prime\prime}(\Phi)\right]^2M
\left[
-\frac{1}{32\pi^2\epsilon}-\frac{1}{32\pi^2}\ln\frac{M^2}{\mu^2}\right].
\end{eqnarray}
and
\begin{eqnarray}
\label{g2c}
\Gamma_{b,{\rm C}}^{(2)}&=&-\frac{1}{8}\int d^5z \, V^{(IV)}(\Phi)\frac{M^2}{8\pi^2}.
\end{eqnarray}

For the sake of concreteness, we consider the classical potential $V(\Phi)=m\Phi^2/2+\lambda\Phi^{4}_{*}/4!$. Using our previous results, the K\"ahlerian effective potential in the noncommutative case can be cast as
\begin{eqnarray}
\label{veff}
K(\Phi)&=&\frac{1}{48}\Big\{ 24m\Phi^2+2\lambda\Phi^4+\frac{3}{\pi}\left(m+\frac{\lambda}{2}\Phi^2\right)^2 \nonumber\\
&&-\frac{\lambda}{16\pi^2}
\frac{1}{\left(m+\frac{\lambda}{2}\Phi^2\right)^2} 
\left[ \frac{1}{\Theta^2}+8\left(m+\frac{\lambda}{2}\Phi^2\right)^4 \right]
\nonumber\\
&&+\frac{3\lambda^2}{128\pi^2}\left(m+\frac{\lambda}{2}\Phi^2\right)\Phi^2
\left[\frac{3}{\Theta^2\left(m+\frac{\lambda}{2}\Phi^2\right)^4}-8\ln\frac{\left(m+\frac{\lambda}{2}\Phi^2\right)^2}{\mu^2}\right]\Big\},
\end{eqnarray}
\noindent
where we used the minimal subtraction scheme to renormalize the theory. It is clear that this expression for $m=0$ displays a singularity at $\lambda=0$, therefore in this case the perturbative expansion is broken, which is a consequence of the noncommutativity in this theory (see also~\cite{WZNC} for the four-dimensional analog of this problem).

Up to now, we have considered only the K\"ahlerian effective action. 
Let us describe the general procedure to obtain the one-loop effective potential taking into account the supercovariant derivatives of the background superfield. As we have already noticed, the one-loop effective action (\ref{g12}) reads
\begin{eqnarray}
\Gamma^{(1)}=\frac{i}{2}{\rm Tr}\ln(D^2+\Psi),
\end{eqnarray}

\noindent
up to a normalization, where $\Psi\equiv -V^{\prime\prime}(\Phi)$. Using the Schwinger representation, we can write this effective action as
\begin{eqnarray}\label{eq:1}
\Gamma^{(1)}=\frac{i}{2}\int d^5z\int\frac{ds}{s}e^{is(D^2+\Psi)}\delta^5(z-z')|_{z=z'}.
\end{eqnarray}

\noindent
We then introduce the operator
\begin{eqnarray}
\label{expo}
\Omega(s)=e^{is(D^2+\Psi)}\,,
\end{eqnarray}

\noindent
which can be expanded in a power series in the supercovariant derivatives as
\begin{eqnarray}\label{Oexp}
\Omega(s)=1+c_0(s)+c_1^{\alpha}(s)D_{\alpha}+c_2(s)D^2~.
\end{eqnarray}

\noindent
We note that higher degrees of the spinor derivatives can be reduced to the structures  which are already present in Eq.~(\ref{Oexp}) by using the rules $D_{\alpha}D_{\beta}=i\partial_{\alpha\beta}-C_{\alpha\beta}D^2$, $(D^2)^2=\Box$ and $D_{\alpha}D^2=-i\partial_{\alpha\beta}D^{\beta}$. The coefficient functions $c_0,c_1,c_2$ depend analytically on $s$, the superfield $\Psi$ and its supercovariant derivatives, and the space-time derivatives $\partial_{\alpha\beta}$, which act on the delta function appearing in Eq.~(\ref{eq:1}).

The operator $\Omega(s)$ satisfies the differential equation
\begin{eqnarray}
\frac{1}{i}\frac{d\Omega}{ds}=\Omega(D^2+\Psi)\,.
\end{eqnarray}

\noindent
Substituting here the explicit form for $\Omega(s)$ in Eq.~(\ref{expo}), we obtain a coupled set of differential equations for the coefficient functions $c_0,c_1,c_2$,
\begin{subequations}
\label{sys}
\begin{align}
\frac{1}{i}\frac{dc_0}{ds}&=c_0\Psi+c_2(\Box+D^2\Psi)+c_1^{\alpha}(D_{\alpha}\Psi)+\Psi,\\
\frac{1}{i}\frac{dc_1^{\alpha}}{ds}&=-ic_{1\gamma}\partial^{\gamma\alpha}+c_1^{\alpha}\Psi+c_2D^{\alpha}\Psi,\\
\frac{1}{i}\frac{dc_2}{ds}&=c_0+c_2\Psi+1.
\end{align}
\end{subequations}

\noindent
As $\Omega(s=0)=1$, the initial conditions are $c_0(0)=c_1^{\alpha}(0)=c_2(0)=0$. Since this is a linear inhomogeneous system of differential equations, the solution is of the form $c_i(s)=b_ie^{i\omega s}+d_i$, where $b_i$ and $d_i$ are some $s$-independent coefficients. Substituting this ansatz into the equations (\ref{sys}), one finds for the solution of the homogeneous equation,
\begin{subequations}
\label{sys1}
\begin{align}
(\omega-\Psi)b_0&=b_2(\Box+D^2\Psi)+b_1^{\alpha}(D_{\alpha}\Psi),\\
(\omega-\Psi)b_1^{\alpha}&=-ib_{1\beta}\partial^{\beta\alpha}+b_2D^{\alpha}\Psi,\\
(\omega-\Psi)b_2&=b_0,
\end{align} 
\end{subequations}

\noindent
and for the particular solution of the inhomogeneous one,
\begin{subequations}
\begin{align}
&d_0\Psi+d_2(\Box+D^2\Psi)+d_1^{\alpha}(D_{\alpha}\Psi)+\Psi=0,\\
&-id_{1\gamma}\partial^{\gamma\alpha}+d_1^{\alpha}\Psi+d_2D^{\alpha}\Psi=0,\\
&d_0+d_2\Psi+1=0.
\end{align}
\end{subequations}

\noindent
Equations~(\ref{sys1}), after some simplifications, imply in the following equation,
\begin{equation}
b_{1\gamma}\left[ \left( \omega-\Psi+\frac{1}{2}\frac{D^{\beta}\Psi D_{\beta}\Psi}{(\omega-\Psi)^2-\Box-(D^2\Psi)}\right) C^{\alpha\gamma}-i\partial^{\alpha\gamma}
\right]=0.
\end{equation}

\noindent
Since $b_{1\gamma}\neq 0$ (otherwise the solution is trivial), the $\omega$'s can be found requiring that the $2\times 2$ matrix $\Delta^{\gamma\alpha}$ defined as
\begin{equation}
\Delta^{\gamma\alpha}=\left(\omega-\Psi+\frac{1}{2}\frac{D^{\beta}\Psi D_{\beta}\Psi}{(\omega-\Psi)^2-\Box-(D^2\Psi)}\right)C^{\alpha\gamma}-i\partial^{\alpha\gamma}
\end{equation}
must have zero determinant. This condition is solvable in principle, but we will not pursue this solution here. As in four dimensions~\cite{BK0,WZ}, the evaluation of the non-K\"ahlerian part of the effective action can be done using these methods, but it is technically quite difficult and the results, when found, would be extremely complicated. 

In summary we developed a superfield method for calculation of the effective potential in three-dimensional supersymmetric field theories. We succeeded to obtain explicit expressions for the K\"ahlerian effective potential (which depends on superfield $\Phi$ but not on its derivatives) up to two loops, in the noncommutative case; the corresponding results for commutative theories follows from simple modifications in our formulas. In principle, our approach can be directly generalized for higher loops. We have also shown the approach for the much more difficult calculation of the non-K\"ahlerian contributions to the effective action and potential.

\vspace{1cm}

{\bf Acknowledgments.}
This work was partially supported by Funda\c c\~ao de Amparo \`a
Pesquisa do  Estado de S\~ao Paulo (FAPESP), Conselho Nacional de
Desenvolvimento Cient\'\i fico e Tecnol\'ogico (CNPq) and Coordena\c c\~ao de Aperfei\c coamento de Pessoal de N\'\i vel Superior (CAPES: AUX-PE-PROCAD 579/2008). A.C.L. is supported by FAPESP project No. 2007/08604-1.

\end{document}